\documentclass{article}

\usepackage{arxiv}

\usepackage[utf8]{inputenc} 
\usepackage[T1]{fontenc}    
\usepackage{hyperref}       
\usepackage{url}            
\usepackage{booktabs}       
\usepackage{amsfonts}       
\usepackage{nicefrac}       
\usepackage{microtype}      
\usepackage{lipsum}
\usepackage{cite}
\usepackage{cite}
\usepackage{amsmath,amssymb,amsfonts}
\usepackage{algorithmic}
\usepackage{graphicx}
\usepackage{textcomp}
\usepackage{graphicx}
\usepackage{subcaption}
\usepackage{multirow}
\usepackage{mwe}
\usepackage{subcaption}
\usepackage{booktabs}
\usepackage{mathrsfs}

\usepackage{booktabs}

\title{CoMuTe: A Convolutional Neural Network Based Device Free Multiple Target Localization Using CSI}

\author{
  Tahsina Farah Sanam\\
  Department of Electrical and Computer Engineering\\
  Rutgers University\\
  New Brunswick, NJ 108854 \\
  \texttt{tahsina.farah@rutgers.edu} \\
   \And
  Hana Godrich \\
  Department of Electrical and Computer Engineering\\
  Rutgers University\\
  New Brunswick, NJ 108854 \\
  \texttt{godrich@soe.rutgers.edu} \\
}

\begin{document}
\maketitle

\begin{abstract}
With the growth of Internet-of-Things (IoT), Location based Services (LBS) are gaining significant attention over the past years. Location information is one of the important ingredients for many LBS where the system requires to localize multiple targets in indoor setting. As an emerging technique, device-free localization (DFL) is promising to localize the target without attaching any transceivers. In this paper, we propose CoMuTe, the first convolutional neural network (CNN) based device free multiple target localization leveraging the Channel State Information (CSI) from multiple wireless links. The system represents the CSIs as Multi-link Time-Frequency (MLTF) image by organizing them as time-frequency matrices and utilize these MLTF images as the input feature for CNN network. The CoMuTe models multi target localization as a multi label spot classification approach under the assumption that each MLTF image is associated with multiple labels/spots. The localization is performed with a training stage and a localization stage. In the training stage, the CSI based MLTF images are constructed with single target at each location. The constructed images are used to train the CNN via a gradient based optimization algorithm. In the localization stage, the test MLTF image obtained for targets at multiple spots is fed to the CNN network and locations of the targets are calculated using sigmoid activation function in the output layer under the multi label classification framework. Extensive experiments are conducted to select appropriate parameters for the CNN architecture as well as for the system design. The experimental results demonstrate the superior performance of CoMuTe over existing multi target localization approaches.
\end{abstract}


\section{Introduction}\label{sec:introduction}
%
%
%
%
Over the past decade, indoor localization is highly in demand for developing smart home systems. Location information is one of the essential ingredient for many location-based services (LBS). However, most of the existing approaches are device-based, where the targets require to carry electronic devices or tags \cite{Nuzzer,augmented1,augmented2,CSI-MIMO}.
Device-free localization (DFL) is an emerging technique where the targets are not required to be equipped with devices \cite{Sanam_journal,pilot,LiFS,DisLoc}. Therefore, no active participation of the target is required in the localization process. The ability to localize various targets in a device free setting at indoors can potentially support a broad array of applications including elder care, rescue operations, vehicle parking management, building occupancy statistics, security enforcement, etc. \cite{application1}. Hence, with device free localization techniques, existing wireless infrastructures can be empowered with the ability of location awareness while, at the same time, the normal communication tasks can be left undisturbed. 

A wide range of emerging technologies including wireless sensors, radio frequency identification (RFID) and visible lights have been explored in the field of indoor localization\cite{RFID1,RFID2,VL1,VL2}. All of these works depend on complex hardware setup. Recently, indoor localization based on wireless local area networks (WLAN) are getting more popular due to open access and lower cost of different wireless signals \cite{pilot, Xu_mob_com,Sanam1}. In wireless signal-based device free localization, the location of transceiver-free target is estimated by analyzing the change in feature pattern of the wireless signal that is being interfered by the presence of the target in the area of interest \cite{why_IL1,survey1}. One of the most widely used signatures of target locations is Received signal strength (RSS) \cite{whyRSS,DFL_RSS, sae1dcnnRSSIEEEAccess}. DFL utilize the change in feature pattern of RSS to characterize the shadowing effects caused by the target. However, RSS readings are coarse-grained MAC-layer information which is the superposition of indoor multipath components of all wireless links. Due to multipath fading and noises in changing environment, RSS readings vary with time and therefore become unreliable for accurate localization. In order to overcome various drawbacks of RSS, recently fine-grained PHY-layer channel state information (CSI) has been leveraged for target localization \cite{RSS_to_CSI,fuseloc}. In IEEE 802.11n communication, CSIs can be obtained from Multiple Input Multiple Output (MIMO) Orthogonal frequency Division Multiplexing (OFDM) systems. Unlike RSS, each CSI measurement provides us with amplitude and phase information for subcarrier level channels for each antenna link. These fine grained CSI is not only richer in multipath information, but also more stable than RSS for a given location. Therefore, CSI is considered as a preferable choice of wireless signature to realize an improved indoor localization system\cite{survey2}.

The research in indoor localization has been explored with different technologies \cite{DFL_RSS,sae1dcnnRSSIEEEAccess,fuseloc}. A signal dynamics model is proposed in \cite{DFL_RSS} to allow tracking of transceiver-free object. This work relies on RSS value of the wireless sensor network. The work in \cite{sae1dcnnRSSIEEEAccess} combines a Stacked Auto-Encoder (SAE) with a one-dimensional CNN to extract key features from sparse RSS for training the CNN network. A CSI based device free localization is proposed in \cite{fuseloc}, where the location of a target is estimated using information fusion of features extracted from CSI measurements. However, these state-of-the-art methods are focused on localizing a single entity. One of the great challenges for real-world application of indoor localization is to localize multiple targets and multiple target scenario is usual in practical application. For single user case, the site is surveyed by dividing the target area into cells and measuring the CSI data for all cells one by one during the training stage\cite{Xu_mob_com,Sanam_journal}. However, For multiple target case, all the combinations of entities over all calibration locations are required to construct the fingerprints, which grows exponentially with the number of fingerprint locations as well as with the number of targets, making laborious calibration inevitable. 

Different methods have been carried out to overcome the challenges of indoor localization in multi target scenario. In \cite{why_mul3}, a localization model of distance, transmission power and the signal dynamics caused by the objects is proposed, but requires a dense deployment of the sensor network. Therefore, the hardware cost of this approach is very high.  A RFID based method to localize the targets are proposed in LANDMARC\cite{landmark}, but requires a lot of calibration. 
Recently compressive sensing (CS) based techniques has shown good promises in multi target localization. 
In CS-based methods the multi-target problem is generally formulated as a sparse signal reconstruction problem, where the target locations are estimated by taking the advantage of CS theory in sparse recovery. However, CS based positioning algorithms are typically able to achieve good accuracy when
only sparse information is available, thereby limiting the performance when the information is not sparse. A CS-based DFL method FitLoc was introduced in \cite{fitloc} which proves that the restricted isometry property (RIP) is satisfied by model-based dictionary with high probability and estimates the sparse location vector based on the greedy matching pursuit (GMP) algorithm. 
The work in E-HIPA develops an adaptive orthogonal matching pursuit (OMP) algorithm to estimate the number and locations of targets \cite {E_HIPA}. However, all of the above state-of-the-art DFL approaches utilize RSS for multi target localization. A CSI based multi-target DFL is proposed in \cite{SA_M_SBL} under the CS framework which exploits  support knowledge-aided multiple sparse Bayesian learning to utilizes faulty prior information for joint sparse recovery. However, Its accuracy is sensitive to the availability of a prior information of target positions. Another CSI based Multi target DFL is proposed in \cite{CSI_MT2}, where an iterative location vector estimation algorithm is developed under the multitask Bayesian compressive sensing (MBCS) framework. However, the key idea of aforementioned CS-based approaches is to seek appropriate location-dependent CSI features to build a robust relationship between the CSI measurements and target locations. These subjective feature selection as well as dependence upon sparse information limits the robustness of the CS based localization approach.

In this paper we propose a device free multiple target localization leveraging the frequency diversity of Channel State Information (CSI) by formulating the task of localization as a machine learning based classification problem. From  the model design perspective, several works have been reported in literature where localization is modeled as a classification task \cite{KNN,Sanam1}. A KNN based approach is adopted in \cite{KNN}, where as in \cite{Sanam1}, a machine learning based approach is applied for location estimation of a single target using the support vector machine (SVM) based classifier. For the aforementioned methods, professional experiences are needed to tune the model parameters. Also selection of the feature is subjective. To this end, neural networks (NN) based approaches are getting popular which can imitate the signal transition process of neurons and approximates arbitrary math function. Moreover, manual feature selection can be avoided in NN where features are implicitly extracted from input. Therefore, over the past few years, NN has been used for fingerprint based single target localization \cite{deep,BiLoc}. The work in \cite{deep} propose deep learning based indoor fingerprinting system using CSI and use a probabilistic method based on the radial basis function to obtain the estimated location in the localization stage. 
In \cite{BiLoc}, the performance is improved over the \cite{deep} by exploiting bi-modal features of CSI. However, all existing NN based methods are designed to localize single target with fully connected (FC) NN and the complexity increases with the depth of the NN. Moreover, calibration  overhead increases when  the  number  of  fingerprint  locations  increases. So the performance of the model is restricted. To  overcome  the  above  challenges, recently, several approaches leveraging CNN architecture has been reported in literature for indoor localization\cite{restrictedDevice,1Dand2DCnn}. CNN allows to increase the depth of the NN  while keeping the complexity in a proper level.
In \cite{restrictedDevice}, a CNN based NLoS channel classification and ranging error regression models are used leveraging  raw channel impulse response (CIR) information. 
The work in \cite{1Dand2DCnn} develops a layer-wise relevance propagation (LRP) algorithm  to quantify the contribution of the input data to a specific output prediction using a CNN architecture for localization. However, all of these approach are designed to localize a single subject using either RSS or CSI data and left the task of multiple target localization untreated.

To address the challenges of multi target localization, we propose CoMuTe, the first convolutional neural network (CNN) based  device free multiple target localization by exploiting multi label classification framework using CSI based location features. As an important aspect of modern predictive modeling, multilabel classification is rapidly developing and is widely used in different classification purposes \cite{ML2,sigmoid_loss1}. 
In modern classification problems, it often requires to predict multiple labels simultaneously associated with a single instance. Therefore, it represents complex signals that have multiple meanings and helps to capture more information by labeling some basic and hidden patterns. In this respect, multi-label classification is very useful in multi target localization. It addresses the problem of classifying CSI measuremnts with different location labels, when captured for indoor scenario with multiple targets at multiple locations. In CoMuTe, the monitoring area is virtually partitioned into grids of uniform square cells, where each grid is considered as a class or label. Therefore, the multiple target localization problem is formulated as a multi label grid classification problem. To utilize a CNN based multi label classification framework, we transform the CSI measurements from each radio link into a time-frequency feature matrix which corresponds to an image. Thus CSI based time-frequency matrices obtained from different radio links for one location represent different RGB channels of an image. We stack these time-frequency feature matrices for all the links together, and refer this as Multi-link Time-Frequency (MLTF) image. We use this images as the input to the CNN to train the network. 
We implement a multi-label classification approach with the CNN consisting of three convolutional layers and two FC layers. In multi label classification, appropriate choice of loss function as well as the activation function at the output layer allows to train the network on single labels. 
Therefore, the system does not require all the combinations of entities over all calibration locations to construct the fingerprints, which reduces the calibration overhead to a great extent. 

The contribution of the paper can be summarized as follows. Firstly, a novel approach of presenting CSI as MLTF image is proposed. Utilization of MLTF image leads to avoid prepossessing as well as the manual subjective feature selection while helps to utilize the information contained in CSIs from multiple wireless links comprehensively. Secondly, to the best of our knowledge, CoMuTe is the first method that utilizes multi label classification approach with CNN for localizing multiple targets, which analyze each MLTF image and classify it with one or more of the location labels. Lastly, by combining convolutional layers with FC layers, the depth of the NN is extended to improve the accuracy of localization while keeping the complexity in a proper level. Extensive experiments performed in a cluttered indoor environment are used to verify the effectiveness of CoMuTe, demonstrating it outperforms state-of-the-art multi target localization methods. 

The rest of the paper is structured as follows. Preliminaries on the deployment of CoMuTe system as well as the channel state information are described in Section \ref{sec:prelim}. Section \ref{sec:motivation1} presents the motivation behind the proposed system. The CoMuTe system with CNN based multi target localization is introduced in Section \ref{sec:proposedcomute}. Section \ref{sec:Experimental Study} describes the experimental setup and evaluates the
performance of the proposed method. Finally, concluding
remarks are discussed in Section \ref{sec:conclusioncomute}.


\section{Preliminaries}\label{sec:prelim}
\subsection{Deployment of CoMuTe System}
\begin{figure}[t]
\centering
\includegraphics[width=3.3in]{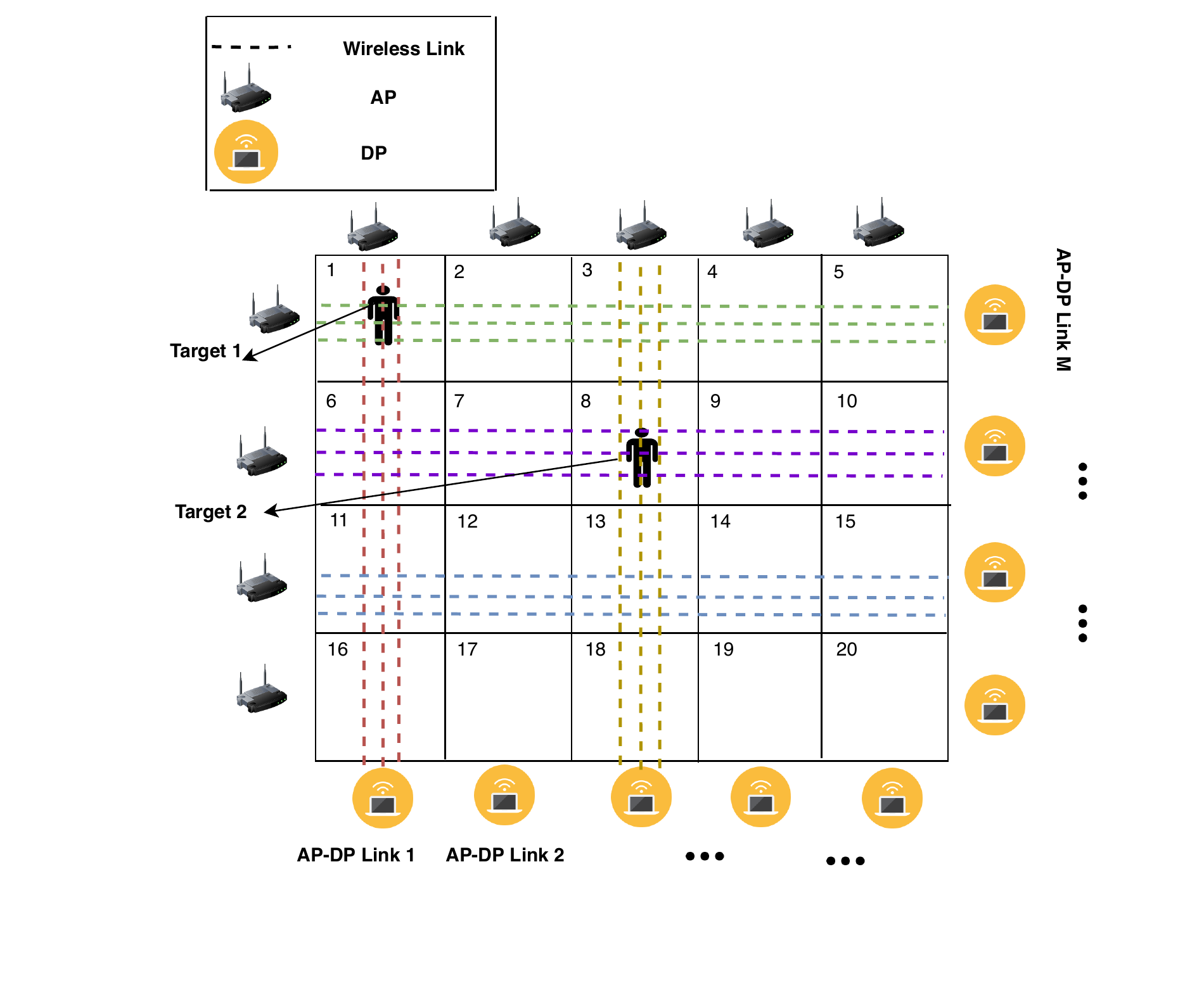}
\caption{System design of device free multi target localization }
\label{system}
\end{figure}

The proposed CoMuTe system consists of three basic hardware elements in a WLAN infrastructure: access points (AP), detecting points (DP) and a server. A WiFi compatible device is used as the DP that interacts with the AP and the server. The sever serves as the processing unit where all the data are stored and processed for localization. TL-WR940N wireless routers are used as the AP. The DPs are equipped with Intel 5300 Network Interface Card (NIC). Each pair of AP and DP establishes a radio frequency (RF) link. Beacon messages are broadcast periodically by the APs. Once the beacon message is received, the DP records the raw PHY layer CSIs across multiple subcarriers from the AP and sends them to the server to store and process. Without loss of generality, the monitoring area is considered as two-dimensional, where multiple transceiver-free targets are randomly distributed over the region as shown in Fig. \ref{system}. 
The monitoring area is divided into N equal-sized grids. M APs, $\{AP_1,AP_2,..AP_m,...,AP_M\}$ are uniformly deployed along half of the perimeter of the monitoring area as the transmitters, and M DPs $\{DP_1,DP_2,...,DP_m,....,DP_M\}$  are deployed along the other half of the perimeter of the monitoring area as the receivers. Each node is placed at the midpoint of a grid side. Thus M AP-DP wireless links are formed by pairs of $\{AP_m,DP_m\}$ transceivers (m $\in$ [1,M]), as shown in Fig. \ref{system}.
Since the targets are spatially distributed over different grids, target locations can be represented as,
\begin{equation}\label{locationvec}
l=[l_1, l_2,....,l_n,...,l_N ]^T,
\end{equation}
where $l$ $\in$ $\mathbb{R}^{N \times 1}$ is the location vector, $l_n$ $\in$ $\{0,1\}$ is the n-th component of l. Formally, multi-label classification is formulated as the problem of finding a model that maps input features to binary output vector by assigning a value of 0 or 1 for each element (label) in the output vector. Therefore, in multi target localization scenario, if there exists a target at grid $n$, $l_n$ is set to 1; otherwise, $l_n$ is set to 0. We denote K as the number of targets. Thus the number of targets \(K=\sum_{n=1}^{N}l_n\). Our goal is to determine the locations of targets using Eq. (\ref{locationvec}) by classifying the CSI measurements that are captured for multiple targets located at multiple locations. 

\subsection{Channel State Information}
One of the fundamental functionality in wireless communication system is to estimate CSI for transmission links, which further facilitates different functionalities such as power control and handover. Training sequence are used for CSI estimation in WiFi. According to IEEE 802.11n \cite{Halperin}, CSIs are estimated during transmit beam-forming procedure by sending sounding PPDUs (physical layer convergence
procedure protocol data unit) from the beamformee to the beamformer. In the time domain, the received signal $r(t)$ is the temporal convolution of transmitted
signal $s(t)$ and channel impulse response $h(t)$:
\begin{equation}\label{Recsig}
r(t) = s(t) * h(t).
\end{equation}
Here $h(t)$ models the comprehensive effects of large scale fading, multi-path fading and shadowing. In frequency domain, the channel matrix can be calculated as
\begin{equation}\label{CFR}
H=R/S,
\end{equation}
where $H$ represents the PHY layer CSIs between the transmitter and the receiver over multiple sub-carriers, $R$ denotes received signal spectrum and $S$ denotes transmitted signal spectrum. 
CSI of a single subcarrier $k$ for one measurement is a complex value \cite{Sanam_journal}, 
\begin{equation}
h_{k}=R_{k}+j I_{k}=|h_{k}|e^{j sin\theta_{k}},
\end{equation}
where $R_{k}$ and $I_{k}$ are the in-phase and quadrature components, respectively; $|h_k|$ is the amplitude, and $\theta_k$ is the phase of $k$-th subcarrier. The amplitude response of subcarrier $k$ is  \(|h_{k}|=\sqrt{R_{k}^2 + I_{k}^2}\), and the phase response is computed by \(\angle h_{k}=\arctan(I_{k}/R_{k})\). However, off-the-shelf commodity WiFI devices are equipped with multiple antennas. For each transmitter-receiver (Tx-RX) antenna pair, $H$ is a $T \times s$ matrix for each AP-DP link, where $s$ denotes the number of subcarriers for each antenna pair and $T$ is the number of measurement packets. Let, there are L number of Tx-Rx antenna pairs in each AP-DP link. CSI measurements are collected at each grid and then the amplitude of CSIs of all Tx-Rx antenna pairs of $m$-th AP-DP link are grouped together to obtain the CSI fingerprint for n-th grid,
\begin{equation}\label{Hall}
\boldsymbol{H}_m^n=[|\boldsymbol{H}_1| |\boldsymbol{H}_2| \ldots |\boldsymbol{H}_l| \ldots |\boldsymbol{H}_L|],
\end{equation}
where m=1,2,...M, n=1,2,...N and $l$ is the index of Tx-Rx antenna pairs for each AP-DP link. Each $|\boldsymbol{H}_l|$ is a $T \times s$ matrix. Therefore, $\boldsymbol{H_m^n}$ $\in$ $\mathbb{R}^{T \times d}$, where $d = s \times L$, the total number of subcarriers from all Tx-Rx antenna pairs. We use the CSI amplitude dynamic measurement to quantify the interference caused by the target as described below. 

\section{Motivation}\label{sec:motivation1}
\subsection{CSI Amplitude Dynamic for DFL}\label{subsec: csidynamics}
When a wireless AP-DP link communicates, the radio signals pass
through the physical area of the network. The transmitted
signal is diffracted, scattered, absorbed or reflected by the targets within the area. Based on wireless communication principles \cite{csidynamics1}, the CSI amplitude (in dB) $\boldsymbol{H_m^n}$ of link m $(1\leq m \leq M)$ when a target locates at grid n $(1\leq n \leq N)$ is given by,
\begin{equation}\label{csidynamics1}
\boldsymbol{H}_m^n= \boldsymbol{T}_m^n - \boldsymbol{R}_m^n - \boldsymbol{D}_{m}^n + \boldsymbol{S}_{m}^n - \boldsymbol{Q}_m
\end{equation}
where $\boldsymbol{T}_m^n$ is the transmission power of link m, $\boldsymbol{R}_m^n$ is the radio
propagation fading of link m, caused by the path loss and related to the
antenna patterns, etc., $\boldsymbol{D}_{m}^n$ is the diffraction fading of the link m, due to a target that is located at grid n which blocks the LOS path of link m (e.g. target 1 blocks LOS path of AP-DP link 1 as shown in Fig. \ref{system}), $\boldsymbol{S}_{m}^n$ is the scattering gain of the link m, due to a target that is located at the NLOS path, (e.g. for target 2  as shown in Fig. \ref{system}, which introduces an additional path which can increase the amplitude gain for link 1), and $\boldsymbol{Q}_m$ is caused by other fading losses such as multipath effect, absorb, etc.

We use the CSI amplitude dynamic measurement to quantify the
interferences caused by the targets in the area of interest. We denote $\boldsymbol{H}_m^n$ as the
CSI amplitude measurement of link m when a target is located at grid n and
$\boldsymbol{H}_m^{amb}$ as the CSI amplitude measurement of link m when a target is located  outside of the monitoring area. From (\ref{csidynamics1}), it can be noted that, $\boldsymbol{T}_m^n$ , $\boldsymbol{R}_m^n$ and $\boldsymbol{Q}_m$  will not change in a dynamic environment \cite{csidynamics1}. Therefore, the CSI amplitude dynamic measurement $\boldsymbol{\hat{H}}_m^n$ received by link m for a target located at grid n can be obtained as,
\begin{equation}\label{csidynamics2}
\boldsymbol{\hat{H}}_m^n=\boldsymbol{H}_m^n - \boldsymbol{H}_m^{amb} = - \boldsymbol{D}_{m}^n + \boldsymbol{S}_{m}^n.
\end{equation}
Moreover, the work in \cite{DFL_RSS} and \cite{csidynamics2} show that the diffraction fading and the scattering gain depend upon the location of a target. Therefore, the CSI amplitude dynamic based device free localization is feasible. 

\subsection{Multi-link Time Frequency Image}\label{subsec: wifiimage}
Once the CSI fingerprints $\boldsymbol{\hat{H}}_m^n$  are obtained using CSI amplitude dynamic for each grid, MLTF images are constructed by transforming CSI fingerprints from all AP-DP links into a feature matrix corresponding to the pixel value of an image. From eq. (\ref{Hall}), it can be noted that, for m-th AP-DP link, $\boldsymbol{H}$ is a $T \times d$ matrix, constructed by grouping T packets for d CSI amplitude dynamic value from all the corresponding TX-RX antenna pairs for one grid. From two dimensional perspective, the elements in the rows are composed by the time samples and for each row, the CSI amplitude dynamic in the columns correspond to subcarriers from all channels. Therefore, for M AP-DP links, we have a $T \times d \times M$ matrix for each location. We refer this matrix as CSI based Multi-link Time-Frequency (MLTF) image. These time-frequency images obtained from multiple AP-DP links reflect different multipath propagation features. Therefore, CSI time-frequency based image obtained from one AP-DP link is considered as one RGB channel of an image. Unlike image, the number of channels for MLTF image depends upon the number of AP-DP links available in the monitoring area. These images collected for the same location are treated as samples for the same class and used to train the CNN based classification model.
\begin{figure}[t]
\centering
\includegraphics[width=3.3in]{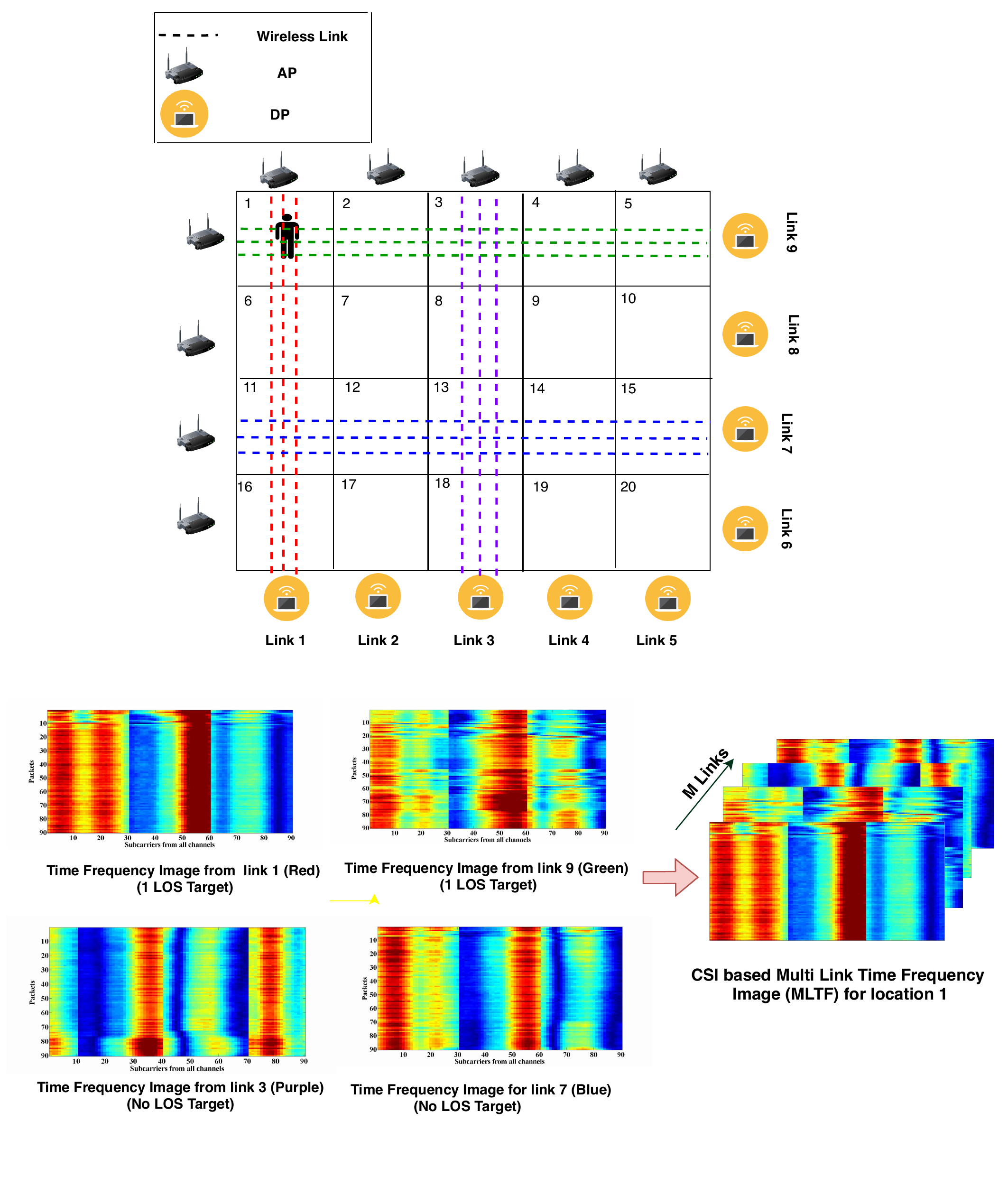}
\caption{Effect of single target on MLTF image for location 1}
\label{motivation1}
\end{figure}

As a fine-grained PHY-layer measurement, CSI captures information on how a signal propagates from the transmitter to the receiver at the subcarrier level. From Section \ref{subsec: csidynamics}, we see that when a target is located at different location, the corresponding CSI amplitude dynamic measurement varies differently. Intuitively, when there are multiple targets in multiple grids, then the corresponding MLTF image should contain some important features induced by individual target's interference on each AP-DP link. Therefore, we assume that, under multi target scenario, each CSI based MLTF image is associated with multiple location labels where the targets are located. Based on this idea, the target positions can be estimated by analyzing the MLTF image through multi label classification approach. 
\begin{figure}[t]
\centering
\includegraphics[width=3.3in]{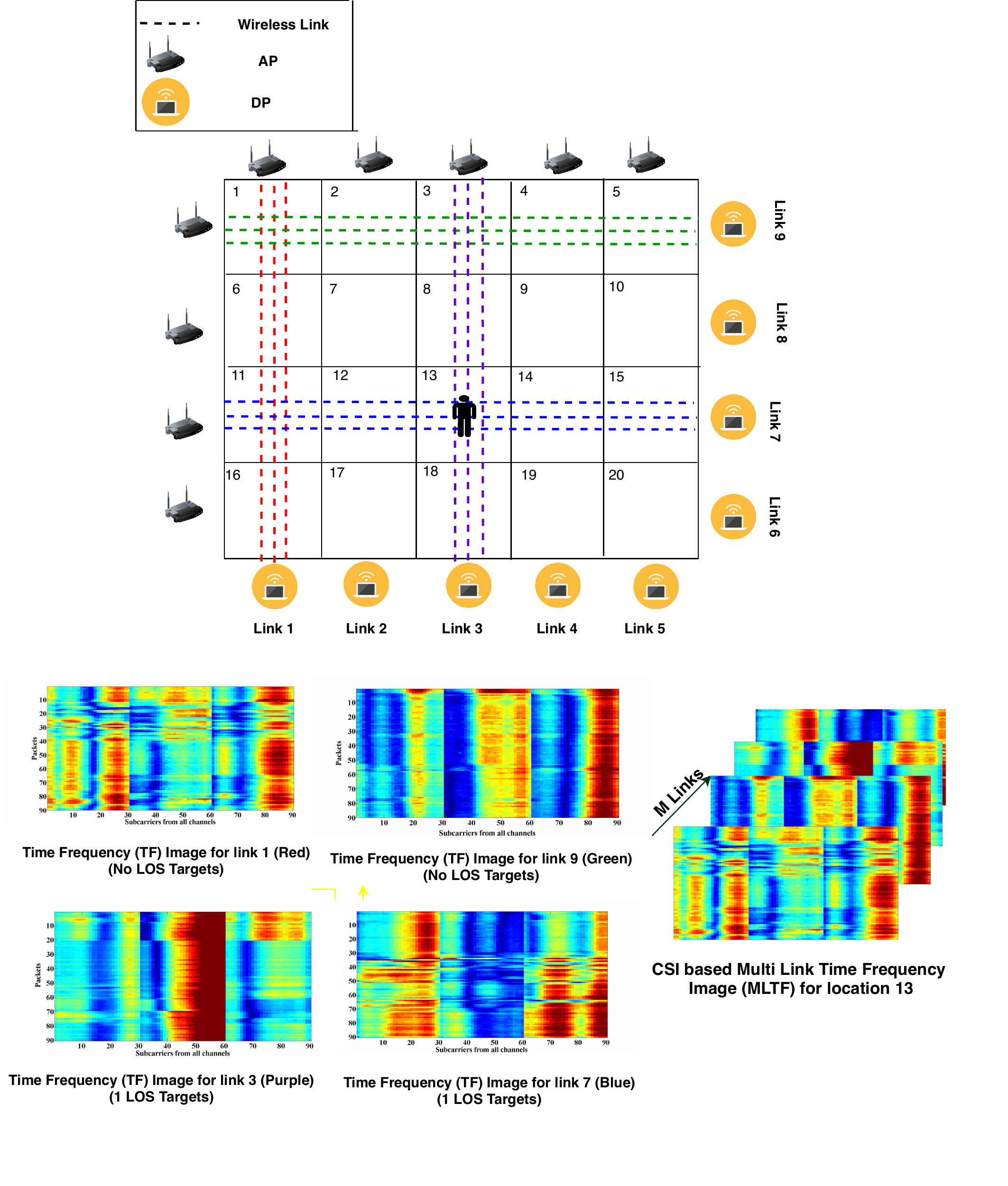}
\caption{Effect of single target on MLTF image for location 13}
\label{motivation2}
\end{figure}
\begin{figure}[t]
\centering
\includegraphics[width=3.3in]{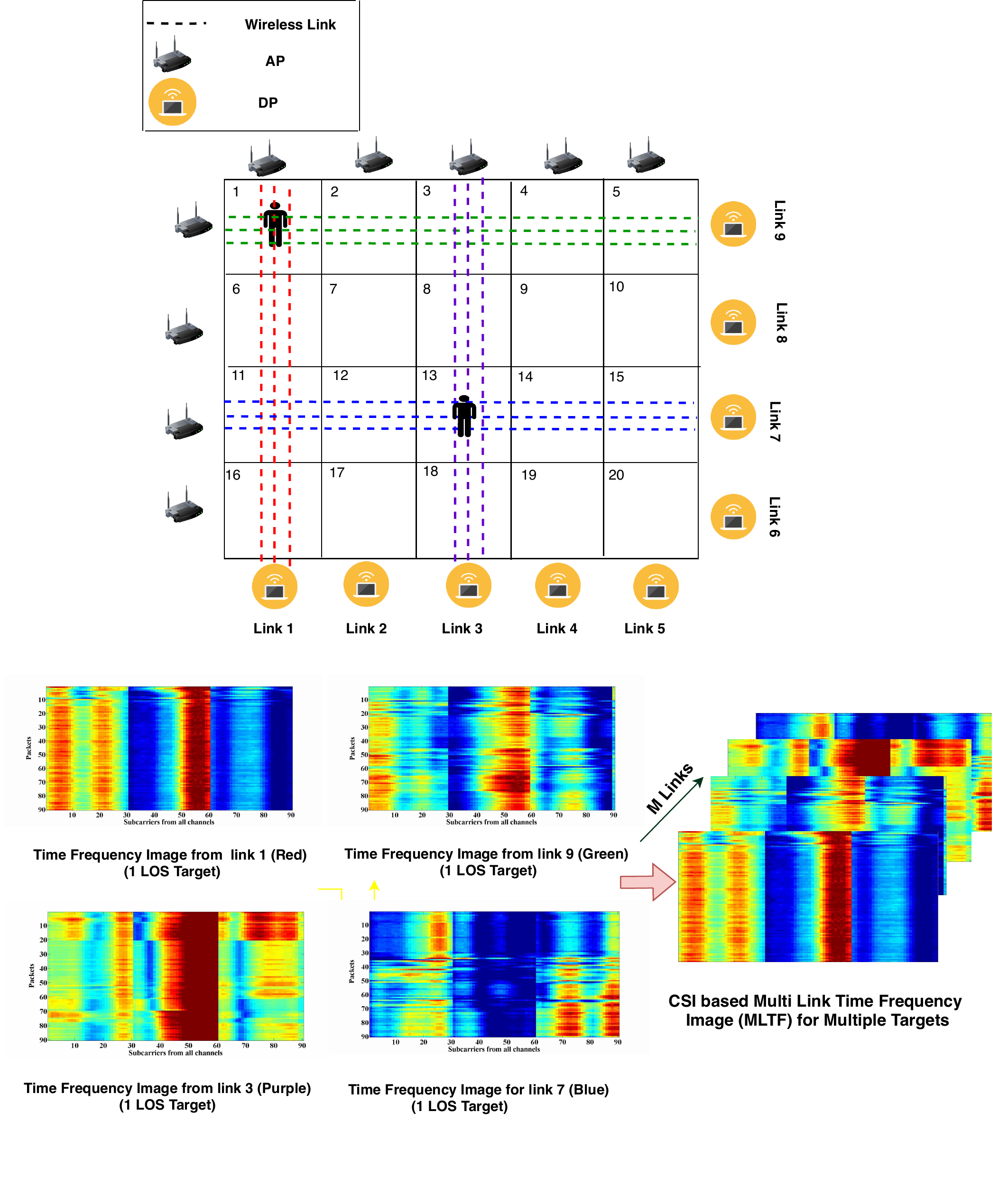}
\caption{Effect of multiple targets on MLTF image}
\label{motivation3}
\end{figure}

To validate the above idea, we conducted some experiments. The results are shown in Fig. \ref{motivation1}, \ref{motivation2}, and  \ref{motivation3}. Fig. \ref{motivation1} illustrates the MLTF image for a target located at grid 1. Due to the presence of one target at grid 1, the LOS communication of AP-DP link 1 and link 9 (red and green links) are affected, while all other links can establish LOS communication. Therefore, the time frequency images of the corresponding links in the constructed MLTF image for grid 1 will reflect certain features that is specific to that grid. Fig. \ref{motivation2} illustrates the MLTF image for one target at grid 13. In this scenario, the LOS communication of AP-DP link 3 and link 7 (purple and blue links) are affected, while all other links can establish LOS communication. In a similar way, the MLTF images constructed for all the other grids with one target in the corresponding grid will have their corresponding location specific unique features. Next, the effect of multiple targets on MLTF image construction is depicted in Fig. \ref{motivation3}. From Fig. \ref{motivation3}, we can see that, two targets are located at grid 1 and 13. For the target at grid 1, the LOS communication of AP-DP link 1 and link 9 (red and green links) are affected. But there is a similarity in color pattern with AP-DP link 1 and link 9 for the one target case of grid 1. For the target at grid 13, the LOS communication of AP-DP link 3 and link 7 (purple and blue links) are affected. Again, there is a similarity in color pattern with AP-DP link 3 and link 7 for One target case of grid 13. Therefore, the corresponding time frequency images in the constructed MLTF image for this multi target scenario will reflect certain features specific to these locations. These features are induced by individual target's interference on each AP-DP link. In CNN based neural network architecture, CNN is able to transform the MLTF image to a set of feature maps, where the discriminative features associated with different locations are extracted by modelling the inter-dependencies among the AP-DP links using the convolutional layers. In addition, more abstract representation of the input MLTF image can be extracted from lower layers to higher layers of hierarchical architecture of CNN. Motivated by this idea, we formulate multi target localization as a multi label classification approach to classify the MLTF image under CNN framework.  

CoMuTe exploits the MLTF image for two reasons. First,
the amplitude dynamic of CSIs depends upon diffraction fading and scattering gain, which are function of a target's location, and therefore a good candidate for designing a device free localization approach. Second, CSI based MLTF image for each AP-DP link (corresponding to one RGB channel of an image) leverages subcarrier information from 3 TX-TX antenna pairs for all the received packets. Therefore MLTF images provide richer time and frequency features for location estimation. In addition, different channels of this image which are constructed from different AP-DP links, reflect different multipath features of the existing wireless propagation system. Therefore MLTF images are rich in both time-frequency as well as spatial features for performing multi target localization.

\section{CNN Based Localization Under Multi-label Classification Framework}\label{sec:proposedcomute}
Multi-label classification problem is one of the supervised learning problems where an instance may be associated with one or multiple labels simultaneously. Currently, Multi-label classification problems have appeared in more and more applications, such as diseases prediction, semantic analysis, object tracking, and image classification, etc. Inspired by the great success from convolutional neural networks (CNN) in single-label image classification in the past few
years \cite{ML2,MLC1}, which demonstrates the effectiveness of
end-to-end frameworks, we explore to learn a CNN based multi-label classification framework for localizing multiple targets in indoor environment. CNN exploits convolutional kernels to extract location depended features in multi target scenario by modelling the inter-dependencies among multiple wirelss links. Use of groups of convolutional layers facilitates CNN to construct increasingly high level representation of the input images at latter layers. Therefore, we apply the  CNN as our model and propose CSI amplitude dynamics based novel MLTF image to apply them as the input to the CNN network for multi-label classification. 
\begin{figure}[t]
\centering
\includegraphics[width=3.4in]{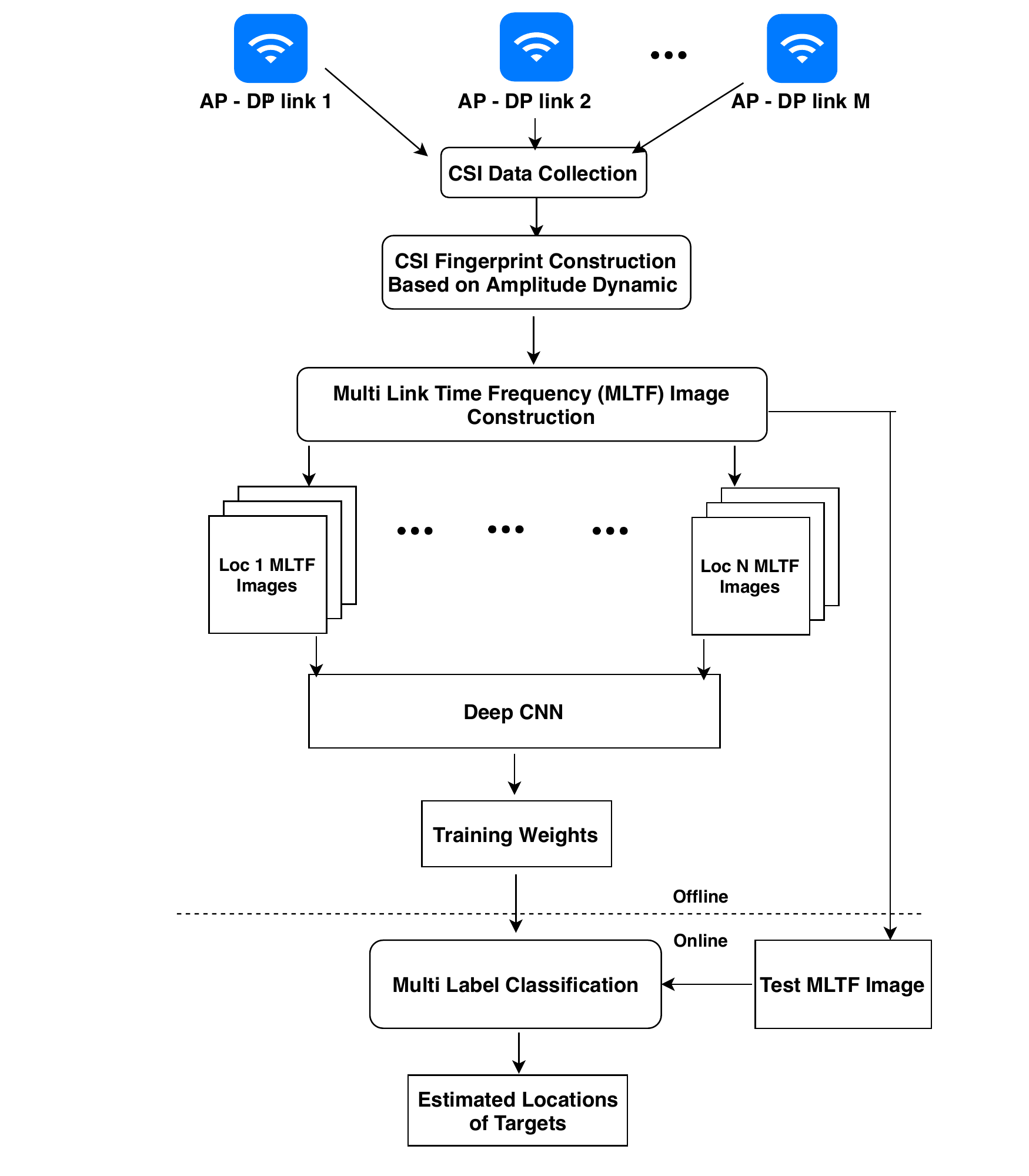}
\caption{The CoMuTe system architecture}
\label{flowchart}
\end{figure}

In the proposed CNN based multi label classification approach, M access points (AP) and M detecting points (DP) are used as WiFi transmitters and receivers, respectively. The DPs are equipped with
the Intel 5300 NIC. With three antennas in the 5300 NIC, each AP-DP link provides CSI readings from 90
subcarriers at the three antennas. Thus, 90 CSI amplitude dynamic for each location under each link can be obtained using (\ref{csidynamics2}). We take 9000 packet samples for
every training location, and construct 100 MLTF images with size
$90 \times 90 \times M$ from all the M AP-DP links. For generalization, we also included multi label instances in our training dataset. Because the number of possible location label combinations was far too large $(2^{N})$ to be exhaustively explored in a reasonable time frame, a subset of location combinations was selected randomly on the criterion that the combinations should be representative of the multi target scenario in a realistic environment. 
The localization is performed in two stages: a training stage and a localization stage. In the training
stage, multiple MLTF images are constructed for every location
and used to train the CNN as in a classical multi-label image classification problem. In traditional fingerprinting
based localization approach, every 
training location requires a separate database, and fingerprints are generated by storing either the measured raw data or learnt features. Unlike the traditional approach, only one group of weights is trained for all the training locations in our CoMuTe system. Therefore the proposed method becomes analogous to a classification or regression problem of classical machine learning. As a result, it can improve the robustness of the system while keeping the amount of stored data much lower. In the
localization stage, when multiple targets are located at different locations, the constructed test MLTF image is fed to the trained neural network and the model outputs the locations of the K targets through the location vector as described in (\ref{locationvec}) using multi-label classification technique.  The overall system architecture of CoMute is illustrated in Fig. \ref{flowchart}. In the next subsections, we  first introduce the structure of the CNN and then present the loss function, optimization, and training and testing methods.

\subsection{Structure of the CNN}
Traditionally, several convolutional and subsampling
layers as well as one or more fully connected layers are incorporated to generate a CNN based deep learning system, which is referred as deep CNN. In CNN, convolutional layers share the same weights between neurons of adjacent layers. Thus local correlations are exploited, which in turn reduce the training time. In addition, more abstract representation of the input image can be extracted from the lower layers to the higher layers of the hierarchical architecture of CNN. Therefore, CNN can extract stronger feature of CSI based MLTF image for indoor multi target localization. The main components of CNN are described in the following.

In CNN, the convolutional layer extracts features from the previous layer’s feature maps through a convolution operation of the input signal with a filter (or kernel), followed by nonlinear activation function. Through the convolution operation, the shift-invariance feature as well as local dependency of input data can be obtained and therefore robust features are extracted. We denote $x_i^j$ as the i-th feature map in j-th layer of the CNN, which is defined as,
\begin{equation}\label{cnnfeature}
x_i^j=\sigma(\sum_{f\in S_{j-1}}w_{i,f}^j*x_i^{j-1} +b_i^j),
\end{equation}
where $x_i^{j-1}$ denotes the i-th feature map in $(j-1)$-th layer, $\sigma(t)$ is a non-linear activation function which is used to avoid obtaining
linear combinations of input data, $S_{j-1}$ is the set of feature
maps in (j-1)-th layer that is connected to the current feature map , $w_{i,f}^j$ is the convolutional kernel to generate the i-th feature map in j-th layer, and $b_i^j$ is the bias of the i-th feature map in layer j. Note that, due to local  weights sharing, $w_{i,f}^j$ is the same for different f. However, in order to ensure that the FC layers have enough number of input features, the input MLTF image is padded and the stride step is set to one in our CoMuTe system. It helps us to avoid reducing the size of the input image by the convolutional layers.



In CoMute, no subsampling or pooling operations are employed to reduce the resolution of the location based feature maps. In literature, besides convolution layers, CNNs very often use pooling layers to reduce data complexity and introduce translation invariant features.
However, this approach is not a strict part of the architecture. There are several examples of CNNs in the time series domain where not every convolutional layer is followed by a subsampling layer or the pooling layer \cite{nopooling}. CoMuTe does not include pooling operations because MLTF image is not similar to actual image. In image classification, the resolution of the feature maps is reduced by the pooling operation  which performs downsampling over a local neighborhood in the feature maps of the previous layer. In this way a lower resolution version of an input image/signal is created that still contains the large or important structural elements, without the fine detail that may not be as useful to the task. In contrast, MLTF image of each link are constructed from the subcarriers of all TX-RX antenna channels for consecutively received packets. Therefore, in the case of CSI based MLTF image, every pixel contains fine descriptions of location features. Application of pooling process will not only confuse these location features but also a lot of useful information will be lost through downsampling.

In order to train the output data of the convolutional layers, CoMuTe utilizes two fully-connected layers that consists of a basic neural network with one hidden layer and one output layer. All of the features obtained from the convolutional layers are concatenated into a single vector, which is then supplied to our fully connected layer. Dropout layers follow each of the densely connected layers with a dropout ratio of 0.6. to prevent feature adaption and avoid overfitting\cite{dropout}. In order to introduce nonlinearity in neural network, we used rectified linear units (RELU) as our nonlinear activation function.
With gradient descent approach, RELU with the non-saturating nonlinearity are much faster in terms of training time than the tanh with the saturating nonlinearities \cite{relu}. Therefore, training time becomes several times faster for CNN with RELUs than their equivalents with tanh units. ReLU can be expressed as follow:
\begin{equation}\label{relu}
f(x) =max(0,x).
\end{equation}
For output layers, we set the number of neurons equal to the number of grids/locations. Therefore, each output neuron corresponds to a grid/location id, which we consider as class. For multi target localization, the final score for each output neuron/class should be independent of each other. The reason is obvious, since targets can be located at any location independent of each other at the time when the corresponding MLTF image was constructed. Thus, softmax activation function can not be used at the output layer. In softmax, the score of each output neuron is converted into probabilities by taking scores of all other neurons into consideration. Therefore, we use the sigmoid activation function on the final layer. Sigmoid converts each score of final node between 0 to 1 independent of what the other scores are. Therefore the output of a neuron can be interpreted as the probability that a target is at the corresponding grid. 
With the sigmoid activation function at the output layer, the neural network models the posterior probability of grid/class $l_n$ as bernoulli distribution:
\begin{equation}\label{sigmoid}
P(l_n|x_i)=l_{n}=\frac{1}{1+e^{-f_n(x_i)}}, n=1,2..,N
\end{equation}
where $l_n$ is output of n-th neuron in the output layer. n is the
index of output neurons while N is total number of output
neurons which is equal to the total number of locations. $x_i$ is the output
of second last layer for the i-th feature map, and $f_n$ means the activation value for $x_i$ and class n. Here the probabilities of each class is independent from the other class probabilities. So we can use the threshold 0.5. If the score for some class is more than 0.5, the MLTF image is classified into that class and $l_n$ is set to 1; otherwise, $l_n$ is set to. There could be multiple classes having a score more than 0.5 independently. Thus the MLTF image could be classified into multiple classes.
\subsection{Loss Function}
The goal of multi-label learning is to predict the label
sets of unseen instances. To achieve this goal, parameters of the feed-forward networks are learned by minimizing some error function defined over the training
examples. These error or loss function measures the difference between
the true location labels and the output data of network. By minimizing
the values of the loss function with the Back propagation (BP) algorithm,
the convolutional weights can be updated with the appropriate optimizer \cite{BP}. The sigmoid (per level) cross-entropy loss has been used for multilabel image classification in MarsNet \cite{sigmoid_loss1} and in MMCNN-MIML \cite{sigmoid_loss2}; therefore, we adopted it in our context. In an indoor environment, targets can be located at multiple locations independent of each other. Therefore, we treat that the location labels are independent, i.e., we assume that the there is no label co-occurrence dependencies for multiple target localization. Under this assumption, we train the network using sigmoid cross-entropy  
as the loss function in order to penalize each output node independently. In other words, we model the output of the network as an independent Bernoulli distributions per label as defined in (\ref{sigmoid}). By utilizing sigmoid cross-entropy as loss function along with the sigmoid activation function at the output layer, an ensemble of single-label binary classifiers is trained, one for each class/location. Each classifier predicts either the membership or the non-membership of one class. The union of all classes that were predicted is taken as the multi-label output. Since in multi target scenario, each MLTF image is associated with multiple labels, we form a label vector $l$ as defined in (\ref{locationvec}), where $l_n$ = 1 means the presence
of a target at grid n and $l_n$ = 0 means absence of a target at grid n for an image. If the ground truth probability for p-th image and n-th class is defined as $\hat{l}_n^p$ , then the following sigmoid cross entropy loss is used for multi-label classification to train the proposed network:
\begin{equation}
  \label{cost}
  \begin{split}
J(w)=-\frac{1}{P}\sum_{p=1}^{P}\frac{1}{N}\sum_{n=1}^{N}[\hat{l}_n^p log(l_n^p) + (1-\hat{l}_n^p)log(1-l_n^p)] \\
\end{split}
\end{equation}
where, 
P is the size
of the training set. The sigmoid cross-entropy in loss function enforces that the output of multiple neurons should be close to one if the targets are located at corresponding locations. 
We train the network to minimize Eq. (\ref{cost}) using appropriate optimization algorithms. 

\subsection{Optimization}
In the training of CNN models, backpropagation algorithm are carried out to train the network until the decease of the loss
function between adjacent iterations falls below a threshold. In CoMuTe, to update the weights during the training phase, we trained our model using Adam \cite{adam} optimizer, which is an optimized version of stochastic gradient descent (SGD). For gradient based optimization, one of the important parameters is the learning rate, $\alpha$. The learning rate controls the speed of adjusting the weights of the CNN network. In SGD a constant learning rate is maintained for every weight update in the network. In contrast, an adaptive learning
rate for each network weight is calculated in Adam optimizer; with the learning rate being adapted as the training progresses. Adam works on adaptive estimates of lower-order moments and computes individual learning rates for different hyper parameter from estimates of first and second moments of the gradients. Adam requires less memory and training time, which is more efficient comparing to stochastic gradient descent optimization (SGD).

\subsection{Localization}
In the localization stage, the MLTF image with multiple targets is fed into the model. If there are N grids in the monitoring area, the model outputs a vector $\textit{l}$, where $\textit{l}$ $\in$ $\mathbb{R}^{N \times 1}$. Each element in $\textit{l}$ can be interpreted as the label corresponding to a grid. 
Using sigmoid cross entropy as loss function, sigmoid activation function at the output layer produces a probability for each of our potential labels independent of what the other probabilities are. During training, these probabilities are used to compute the error, while during testing, we round each of these probabilities to 0 or 1 via a threshold (0.5) in order to indicate whether label n should be applied to the test image. In multi target scenario, there could be multiple classes having a score of more than 0.5 independently. Thus the test MLTF image could be classified into multiple classes denoting presence of targets at corresponding locations.

\begin{figure}[t]
\centering
\includegraphics[width=3.3in]{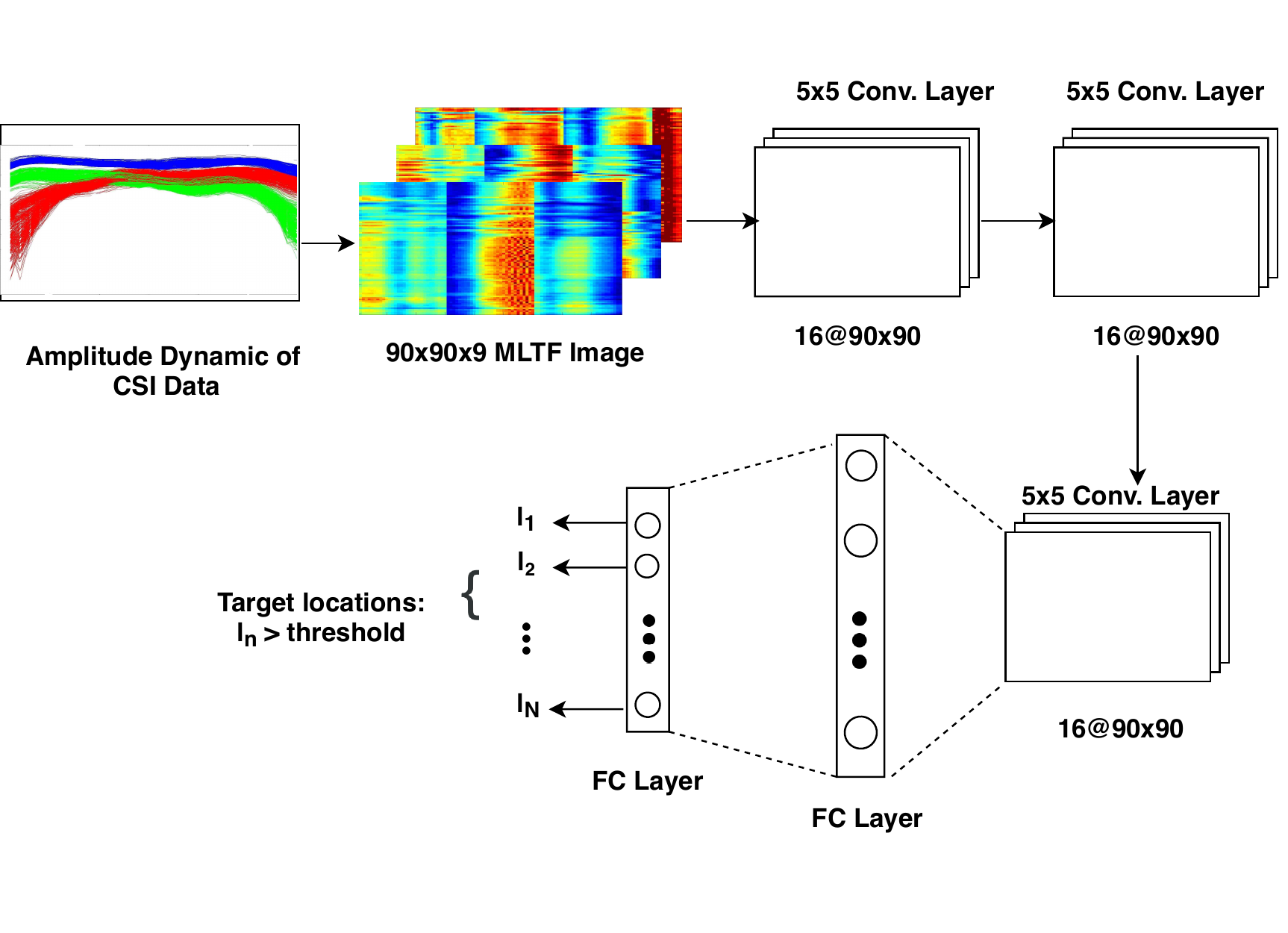}
\caption{CNN architecture for CoMuTe}
\label{CNN}
\end{figure}

Fig. \ref{CNN} illustrates the CNN architecture for training CSI based image data in CoMuTE. To obtain input MLTF images, we first estimate CSI amplitude dynamic value for each of the AP-DP link as in eq. \ref{csidynamics2}. Then, we construct 100 MLTF images for each location, each with size 90 $\times$ 90 $\times$ 9, out of the 9000 received packets. These MLTF images are fed into the CNN network to process
in its convolution and fully connected layers. For each input
image in the first convolutional layer, we set the number
of the convolutional kernel to be 16 with filter size of 5 $\times$ 5. In order to keep the image size unchanged, we use padding in the input image . It helps us to avoid reducing the size of the MLTF image by the convolutional layers. In order to extract the time-frequency information precisely, the stride of the convolutional filter is set to 1. Therefore, we obtain 16 feature maps of size 90 $\times$ 90 at the output of first convolutinal layer. Then, by implementing other two convolutional layers as in Fig. \ref{CNN}, we can obtain 16 feature maps with size 90 $\times$ 90. Next, the output of convolutional layers are flattened as a vector and fed to the fully connected layer. Finally, the output layer uses sigmoid as the activation function to produce the location prediction.

\section{Experimental Study}\label{sec:Experimental Study}
\subsection{Experimental Configuration}

In this section, a series of experiments are conducted
to verify the effectiveness of the CNN based multi-label classification algorithm
for multi target localization. The monitoring area is a $4 m$ $\times$ $5 m$ square region inside a research laboratory in the CoRE Building of Rutgers University. The lab is a cluttered environment, equipped with typical office facilities like desks, shelfs, desktops, chairs etc. TL-WR940N wireless routers are used as the AP. The DP is a mobile device equipped with Intel 5300 Network Interface Card (NIC), which collects the CSI data using the Linux 802.11n tool \cite{Halperin}. For the wireless nodes, we uniformly deploy 9 APs along half of the perimeter of the monitoring area, and 9 DPs along the other half of the area. Each node is placed at the midpoint of a grid side. Thus we establish 9 AP-DP wireless links. Since Intel 5300 NIC is equipped with 3 receiver antennas, we use $3$ TX-RX antenna pairs for each AP-DP link. In OFDM system, each TX-RX antenna pair provides us with $30$ subcarriers. Therefore, for each AP-DP link, we obtain total 90 CSI measurements for each received packet. In order to perform localization in a central processing unit, a host PC (Intel i7-4790CPU 3.60 GHz, 8GB RAM) is used as the centralized server. 

All experiments are conducted during weekdays. The area is virtually partitioned into 20 uniform square grids/cells, each of which is $1m$ $\times$ $1m$ in size. Therefore, the output layer of the CNN has 20 neurons. In the training stage, before targets enter into the area, the CSI packets are received at 0.08s interval and we record for $2$ minutes for m-th AP-DP link.
\begin{table}[t]
  \begin{center}
    \caption{Default values of experimental setup}
    \label{tab:expsetup}
    \vspace{0.3cm}
    \begin{tabular}{l|l}
      \toprule 
      \textbf{Parameters} &  Default values\\
      \midrule 
      Number of grids, N & 20 \\
      Number of links, M & 9 \\
      Number of Targets, K & 5\\
      \bottomrule 
    \end{tabular}\vspace{-0.4cm}
  \end{center}
\end{table}
\begin{table}[t]
  \begin{center}
    \caption{Default values of CNN architecture}
    \label{tab:cnnparam}
    \vspace{0.3cm}
    \begin{tabular}{l|l}
      \toprule 
      \textbf{Parameters} &  Default values\\
      \midrule 
      Input Image size & $90 \times 90$ \\
      Kernel size & $5 \times 5$ \\
      Kernel number & 16\\
      Optimizer & Adam \\
      Learning rate & 0.001 \\
      \bottomrule 
    \end{tabular}\vspace{-0.4cm}
  \end{center}
\end{table}
Then we ask one volunteer person to go through all grids one by one in the area of interest. The CSI packets are received at 0.08s interval and we record for $2$ minutes for m-th AP-DP link for each location. Finally the MLTF image for each location/grid are constructed based on the CSI amplitude dynamic using (\ref{csidynamics2}). To take into account the time domain variation, we conducted $10$ independent measurements on $10$ different days. We collected total 15000 packet samples for every location. The entire dataset is partitioned into training sets, validation sets and test sets using a ratio of 6:2:2. We construct 100 training images for each location with size $90 \times 90 \times 9$ based on the CSI amplitude dynamic, where the third dimension denotes the number of AP-DP links in our experimental setup. For generalization, we also included multi label instances in our training dataset. Because the number of possible location label combinations was far too large $(2^{20})$ to be exhaustively explored in a reasonable time frame, a subset of location combinations was selected randomly on the criterion that the combinations should be representative of the multi target scenario in a realistic environment. Table \ref{tab:expsetup} and \ref{tab:cnnparam} gives the default values of some system parameters and CNN architecture, respectively; that are used in our simulations.

\subsubsection{Compared Approaches} 

We implement two state-of-the-art methods for comparison, i.e., RSS based FitLoc \cite{fitloc}, and CSI based SA-M-SBL \cite{SA_M_SBL}, which are discussed in Section \ref{sec:introduction}. In order to ensure a fair comparison, same experimental setup and parameters are used in all the schemes. Extensive experiments with the above schemes are conducted in a research laboratory to evaluate the performance of the proposed method.

\subsubsection{Evaluation Measures} To evaluate the performance of multi target localization using CNN based multi label classification, we use two kinds of performance metrics. Firstly, we analyze the effect of various parameters on localization performance to identify a good set of
parameter settings for CNN architecture. In order to benchmark the multi label classification performance of the fitted model, we compare our model using the $F1_{\mu}$ measure and Hamming Loss, which are two widely used conventional metrics relevant to multi-label classifications. The $F1_{\mu}$ measure is defined as the harmonic mean of the micro-precision and the micro-recall of all of the labels:
\begin{equation}\label{performmat1}
 \begin{aligned}
		P_{\mu}=\frac{\sum_{n=1}^{N} TP_n}{\sum_{n=1}^{N} (TP_n + FP_n)}\\
		& R_{\mu}=\frac{\sum_{n=1}^{N} TP_n}{\sum_{n=1}^{N} (TP_n + FN_n)}\\
		& F1_{\mu}=\frac{\sum_{n=1}^{N} 2TP_n}{\sum_{n=1}^{N} (2 TP_n + +FP_n + FN_n)},\\
 \end{aligned} 
\end{equation}
where $P_{\mu}$ denotes micro precision, $R_{\mu}$ denotes micro recall, TP is the number of true positive labels, FP is the number of false positive labels, and FN is the number of false negative labels. 
The Hamming Loss operates on each label independently by measuring the ratio of wrongly predicted individual labels to total number of labels over all observed instances:
\begin{equation}\label{locdist}
 \begin{aligned}
		HL=\frac{\sum_{n=1}^{N} FP_n + FN_n}{\sum_{n=1}^{N} (TP_n + TN_n + FP_n +FN_n)},
	\end{aligned} 
\end{equation}
Where TN is the number of true negative labels. Different from the parameter selection part, we also compare the
performance of the proposed method with the state-of-the-art approaches. For this purpose, we use the mean distance error as performance metric, i.e., the average Euclidean distance between the true location and the
estimated location. 
The  experimental results are discussed in the following sections.

\subsection{Analysis of Parameter Setting} 
In order to identify a good set of parameter settings for CNN architecture, we conduct several experiments and analyze the effect of various parameters on localization performance. For this purpose, we use $F1_{\mu}$ and HL for performance evaluation. As described above, we used the validation set to determine when to stop training. After training, the test set is used to test the performance of the trained model. Since we do not use the test set in the training phase, $F1_{\mu}$ and HL on it should be a good approximation of the generalization error of the model. In the experiment, the training sets batch size is set to 256. In addition, the number of epochs is set to 900 to guarantee fairness.

\subsubsection{Impact of Optimizer and Learning Rate}
\begin{figure}[t]
\centering
\includegraphics[width=3.3in]{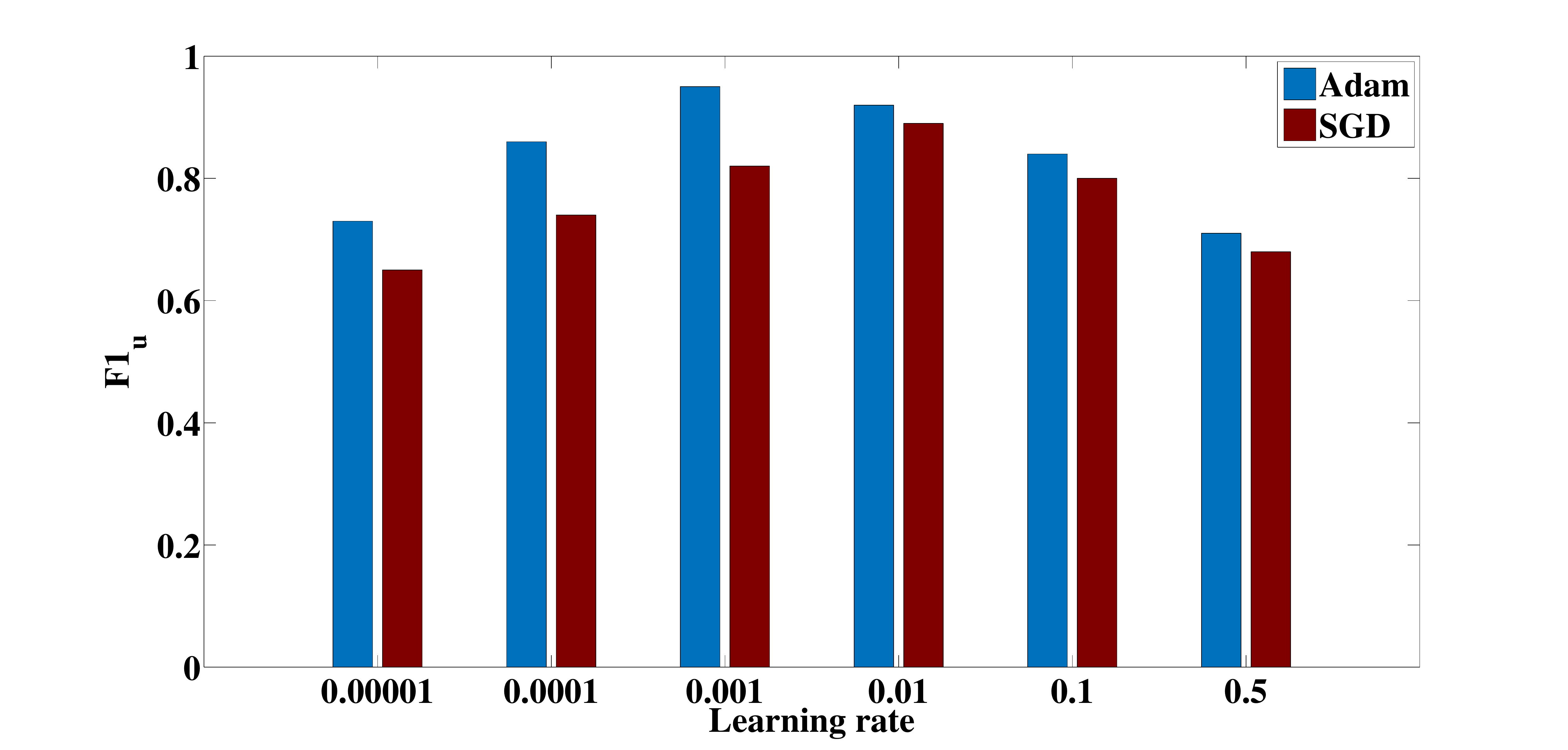}
\caption{$F1_{\mu}$ score for increased learning rate $\alpha$ with Adam and SGD optimizer}
\label{lr}
\end{figure}

One of the important issue in neural network based classification approach is choosing a learning rate and optimizer. In this work, we conduct experiment with two optimizer, SGD and Adam by varying the learning rate, $\alpha$. Fig. \ref{lr} illustrates the $F1_{\mu}$ for different $\alpha$. As the learning rate is increased from 0.0001 to 0.5, CNN model with Adam optimizer obtains the maximum $F1_{\mu}$ score when the initial learning rate is 0.001. For SGD optimizer, the maximum $F1_{\mu}$ score is obtained when the learning rate is set to 0.01. Basically, when the learning rate is low, it may not allow the CNN to converge within 900 epochs. On the other hand, when the leaning rate is high, the BP algorithm diverges or hops back and forth over the valley repeatedly. As a result the network could not reach the best convergence. It can be also noticed that, Adam optimizer outperforms SGD for all the learning rates. This is because, unlike SGD, Adam updates the learning rate for each network weight; with the learning rate being adapted as the training progresses \cite{adam}. Adam requires less memory and training time, which is more efficient comparing to SGD. Therefore, in this study we chose Adam optimizer with the standard default parameters and an initial learning rate of 0.001 to train the network.

\subsection{Performance Comparison with Existing methods} 
Once the required parameters to model the CNN are selected to obtain the best performance, we compare our proposed method with state-of-the-art approaches for multi target localization. We evaluate the performance in terms of mean distance error by considering the following parameters: the number of targets: K, the number of AP-DP links: M, and the cell resolution with width w.
\subsubsection{Performance Comparison for Different Number of Targets} 
\begin{figure}[t]
\centering
\includegraphics[width=3.3in]{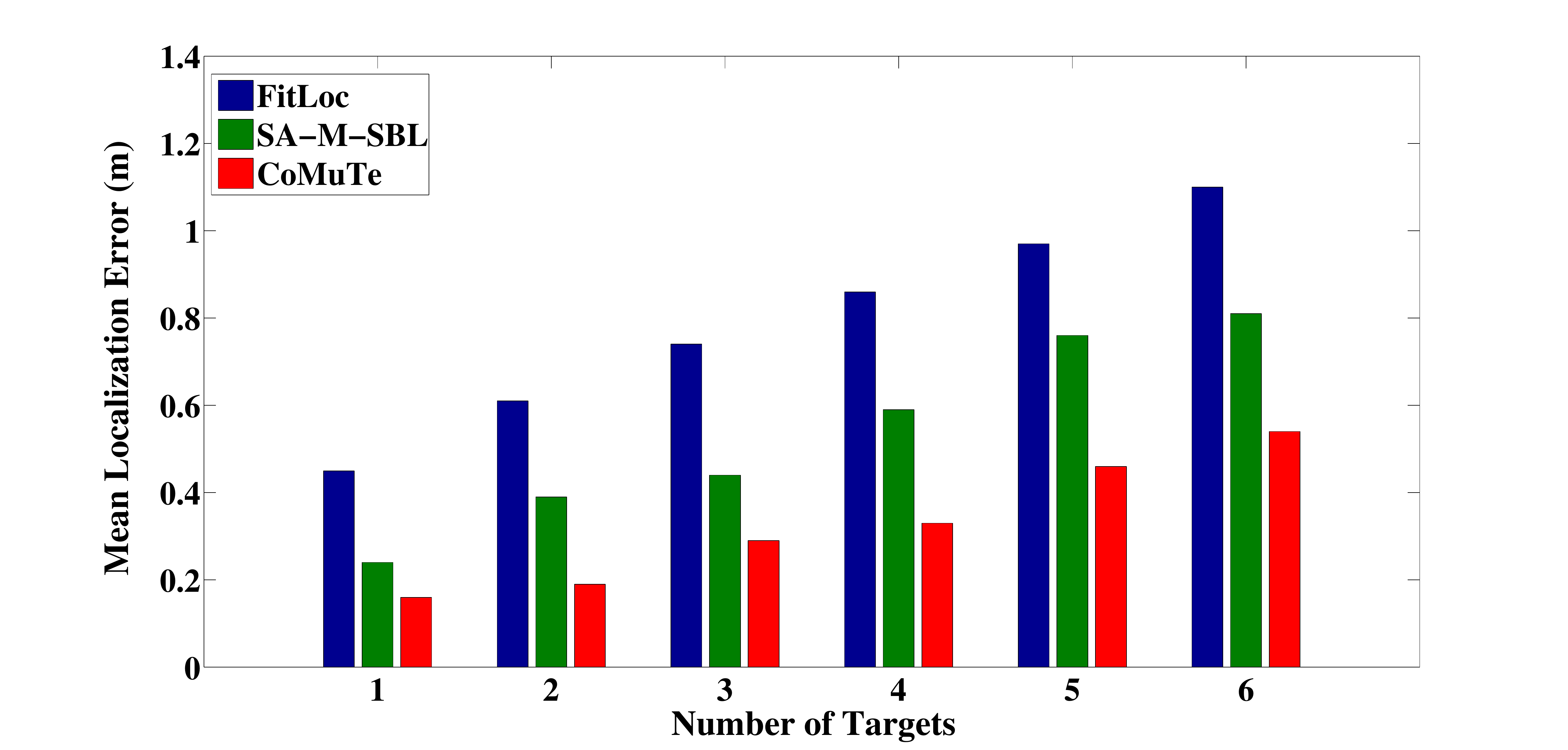}
\caption{Impact of different number of targets}
\label{target}
\end{figure}

In this section, we investigate the performance of our method for different number of targets. 
Precisely, we investigate for how many targets $K$, CoMuTe can obtain a mean distance error that is less than 0.50 m (which is half of the cell width). Fig. \ref{target} plots the mean localization error for three representing methods as a function of number of targets, $K$. As expected, localization error is an increasing function of $K$. As we can see, the proposed method with multi label classification approach is more accurate than other CS based methods. When the number of targets increases from 1 to 6, CoMuTE can localize 5 targets with mean localization error of 0.46 m. The corresponding localization errors of FitLoc and SA-M-SBL are much bigger than CoMuTE. Basically, the sparsity level of the unknown location vector increases with the increase of the number of targets. Hence, the reconstruction accuracy of the location vector will be degraded according to the principle of the CS theory. Therefore, the localization accuracy of all CS-based methods decrease. In contrast, CNN based multi label classification approach does not depend upon sparse recovery and therefore provides us with the lowest localization error among all the other methods when the number of targets increases. However, for 6 targets, the mean localization error in CoMuTe increases to 0.54 m, which is more than the half of the cell size. Assuming that the target is located at the center of a cell, this result indicates that the targets are localized with an average error of 0.54 m from the center of the cell. Therefore we chose 5 targets as our default value for this experiment.

\subsubsection{Performance Comparison for Different Link Number} 
\begin{figure}[t]
\centering
\includegraphics[width=3.3in]{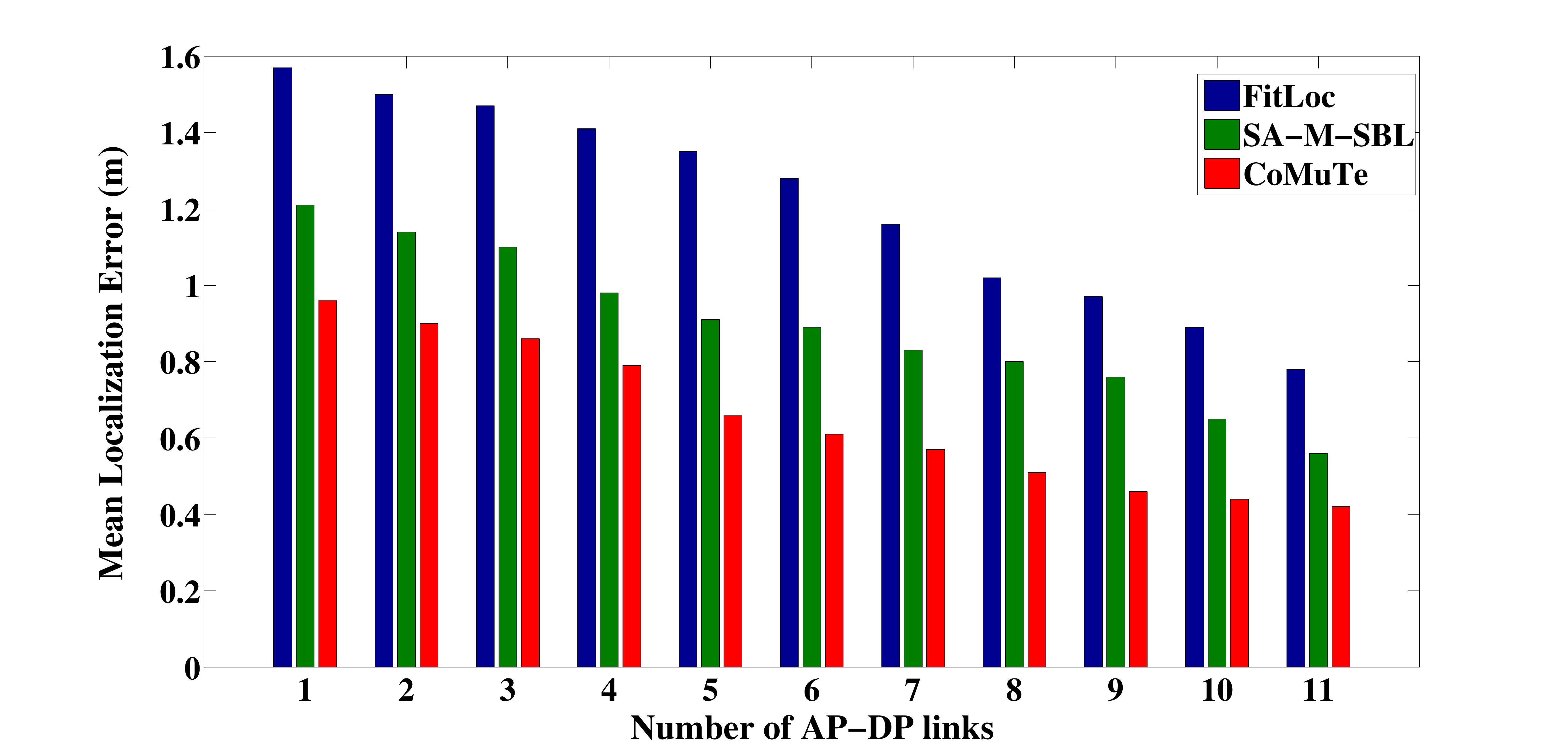}
\caption{Impact of different number of AP-DP links}
\label{link}
\end{figure}

In this subsection, we turn our attention to the impact of
the number of AP-DP links, $M$, on the localization performance of the
proposed method. In this experiment, we increase the number of AP-DP links from 1 to 12. Fig. \ref{link} shows the average localization errors for different number of links, M. It can be seen that the mean localization errors for all methods decrease when the link number increases. This trend is consistent with the fact that the increase in the number of links provides us with more useful information for the target localization. In addition,
benefiting from the theory of CNN, CoMuTe outperforms the state-of-the-art approaches even for a lower number of links. In CoMuTe, each sub-image for one AP-DP link in the MLTF image corresponds to one RGB channel of an actual image. Therefore, different channels of MLTF image which are constructed from different AP-DP links, reflect different multipath features of the existing wireless propagation system. Therefore MLTF images are rich in both time-frequency as well as spatial features. Consequently, CNN can extract better feature maps for each location through the convolution operation to perform multi label classification. This in turn results in a better localization accuracy for multi target localization. However, we see that the localization error for CoMuTe does not decrease significantly when number of links increases to 10 or more. Therefore, in this experiment we chose 9 AP-DP links to balance between the localization performance and deployment cost.

\subsubsection{Performance Comparison for Different Cell Resolutions}
\begin{figure}[t]
\centering
\includegraphics[width=3.3in]{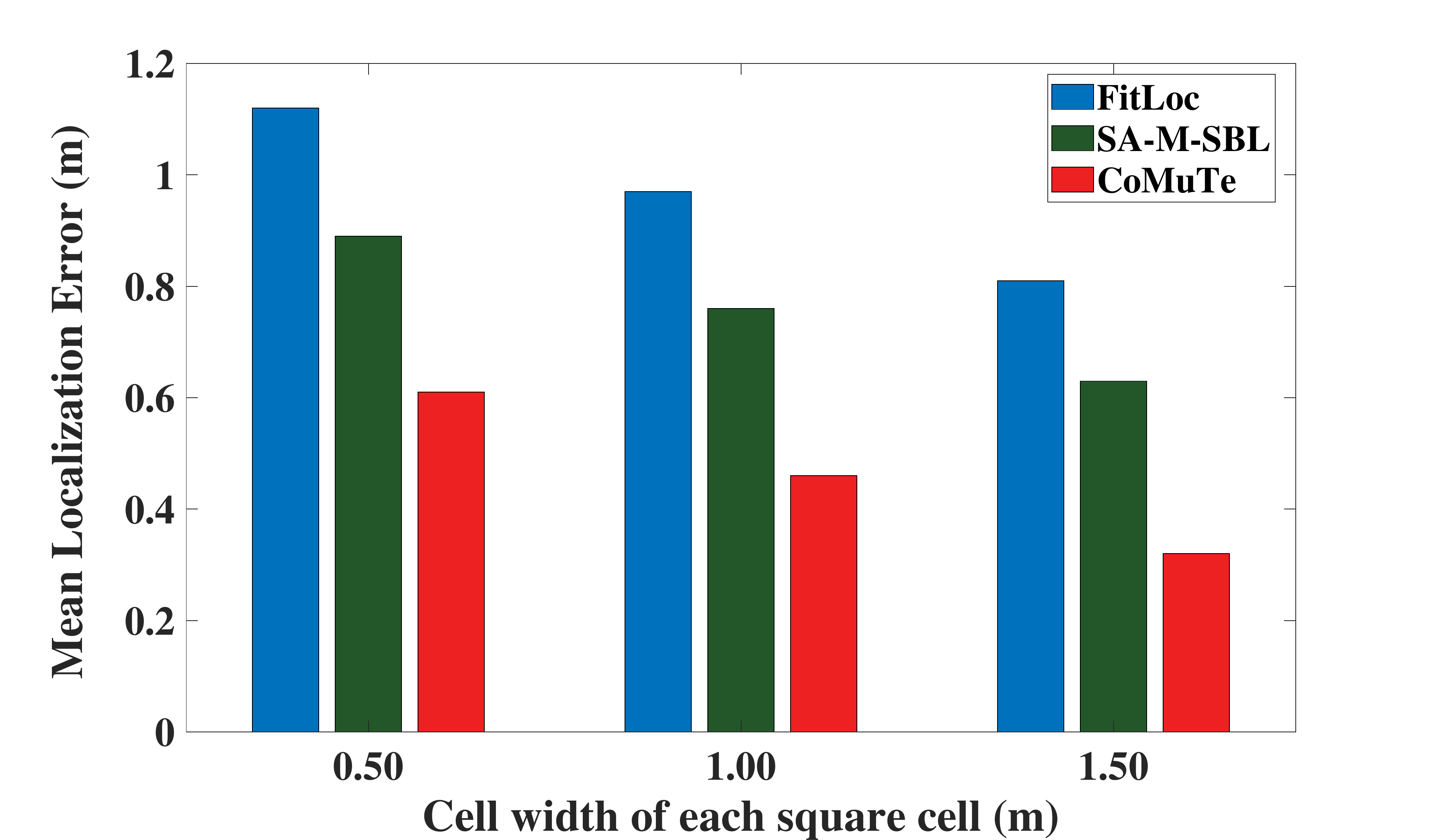}
\caption{Impact of different cell width}
\label{cell}
\end{figure}

To evaluate the effect of cell resolution on localization performance, we performed the experiment with three dataset, each with a different cell resolution. We considered three different cell widths for the square cells, which are 0.5 m, 1.00 m, 1.5 m, respectively. We assume that the targets are located at the center of the cell. Therefore, cell resolution effectively imply the gap between adjacent targets for the underlying system design. Fig. \ref{cell} presents the mean localization errors for different cell widths. 
It can be seen that the cell resolution has a significant impact on the localization performance. When the cell width becomes small,
CSI data collected from the training locations does not vary widely. As a result, it is difficult for the CNN network to make a precise prediction. Therefore, with the decrease in cell width, the mean distance error increases. On the other hand, if the cell width increases, then mean distance error decreases further. Note that, with the increase in cell width, the total number of training locations gets smaller. Therefore, we chose the cell width of 1.00m to balance between the accuracy and built in error. 
\subsubsection{Effectiveness of CoMuTe}
\begin{figure}[t]
\centering
\includegraphics[width=3.3in]{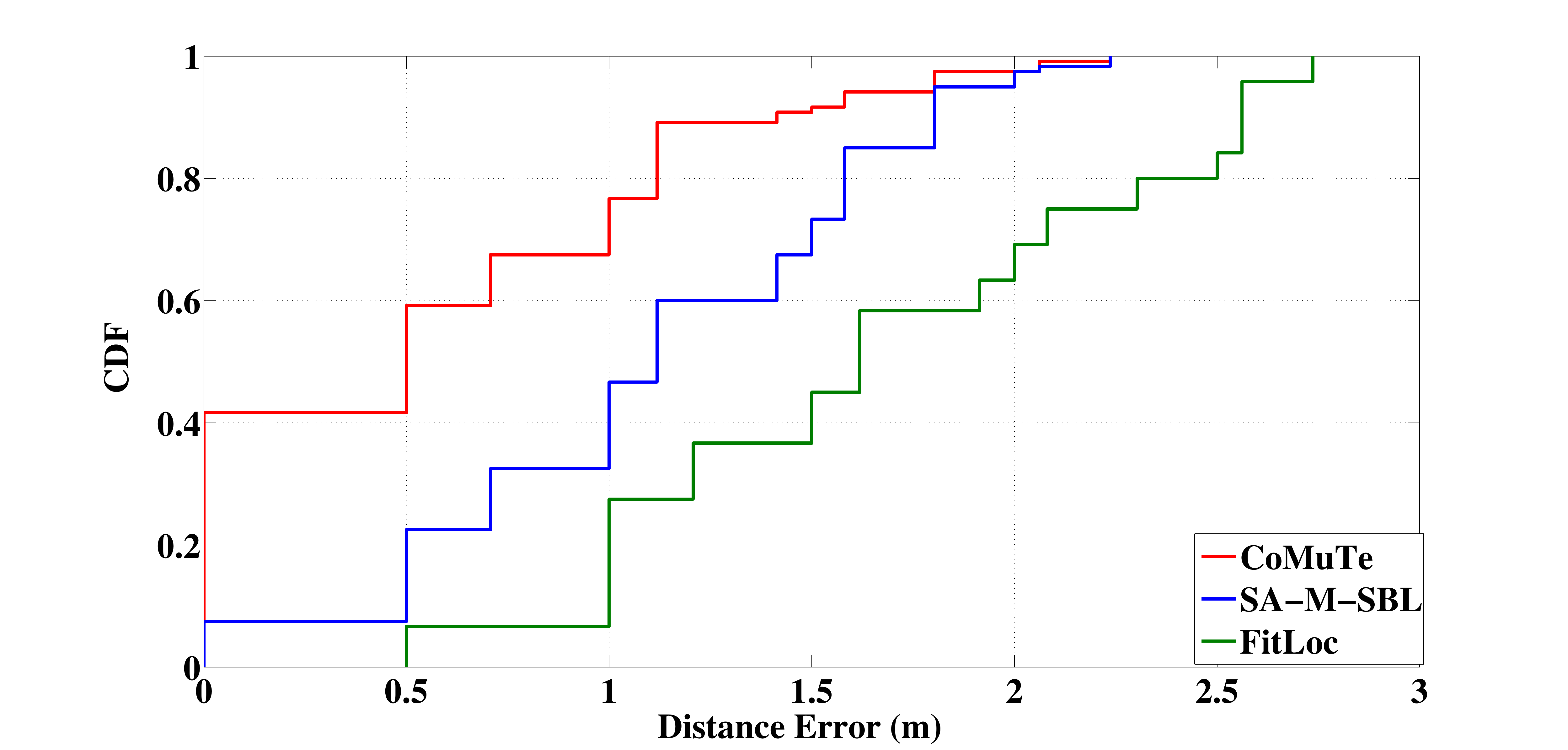}
\caption{CDF of localization error}
\label{cdf}
\end{figure}

FInally, we attempt to discuss the effectiveness of CoMuTe in terms of CDF of localization error.
Fig. \ref{cdf} illustrates the CDF of localization error for all the methods. As we can see, CoMuTe has approximately 42\% of the test examples having an error less than or equal to 0.5m, while that for the other methods is 9\% or less. We also find that approximately 68\% of the test examples for CoMuTe have an error under 1 m, while the percentage of test examples having an error smaller than 1 m are 32\%, and 8\% for SA-M-SBL and FitLoc, respectively. Thus, CoMuTe achieves the best performance in terms of distance error for multi target localization.

\section{Conclusion}\label{sec:conclusioncomute}
This paper presents CoMuTe, the first convolutional
neural network based multi target localization using multi label classification approach. The
algorithm is applied in CSI-based device free localization with multiple targets, where CSI information for all the subcarriers are collected from multiple APs. CoMuTe represents these CSIs as a MLTF image by organizing them as time-frequency matrices for each link. Stacking the time-frequency matrices from all wireless links, CoMuTE utilizes these MLTF images as the input of the CNN model. A five layer CNN with three convolutional layers and two fully connected layers is utilized to extract the location specific features from the MLTF images. By exploiting sigmoid cross entropy as loss function and sigmoid activation function at the output layer, the model internally creates several models, one for each location. Finally the CNN based deep learning model predicts the probability for each location independently trough a multi label classification approach. Extensive experiments are conducted to select appropriate parameters for the CNN architecture. Performance were evaluated for different test cases, such as number of targets, number of wireless links and various cell resolutions. Experimental results demonstrate that the CoMuTe outperforms the state-of-the-art RSS and CSI based multi-target DFL methods with centimeter range accuracy.


\bibliographystyle{unsrt}  
\bibliography{references}  






\end{document}